\documentstyle[12pt,A4,epsf]{article}
\begin{document}
\begin{center}
{\bf \large

Meson Production and Bose-Einstein Pion Correlations 
          in High Energy Collisions}
\end{center}
\normalsize
\medskip
\begin{center}
F. Cannata,$^{a,b,}$\footnote{E-mail: cannata@bologna.infn.it }
J-P. Dedonder,$^{a,c,}$\footnote{E-mail: dedonder@paris7.jussieu.fr }
and M. P. Locher$^{a,}$\footnote{E-mail: locher.@psi.ch} \\
\end{center}
\small
$^a$ {\it Paul Scherrer Institute, CH-5232 Villigen PSI, Switzerland\\}
$^b$ {\it Dipartimento di Fisica and INFN,
  I-40126 Bologna, Italy \\}
$^c$ {\it Laboratoire de Physique Nucl\'{e}aire, Universit\'{e} Paris 7 -
Denis Diderot, F-75251 Paris Cedex 05 and Division de
Physique Th\'{e}orique, IPN
F-91406 Orsay, France \footnote{The Division de
Physique Th\'{e}orique is a Research Unit of the
Universities of Paris 11 and 6 associated to CNRS.
 }}\\

\vspace*{30mm}
\normalsize
\centerline{{\bf Abstract}}
\noindent
The pion Bose-Einstein correlation parameters determined by OPAL show a weak
dependence on pion multiplicity. We argue that the observed increase of the 
interaction radius, and the decrease of the chaoticity parameter could be
due to an increasing fraction of other mesons at high multiplicities.

\vspace*{15mm}
 
\noindent
\newpage
 
  Pion pair correlations\cite{Boa} have been studied from high to low energies,
some recent papers from which earlier references can be traced are
\cite{Ale} to \cite{Ama2}.
The correlation function $C_2$ is
\begin{equation}
          C_2(p,q) = W_2(p,q)/W_1(p)W_1(q) - 1
\end{equation}
where $p$ and $q$ are the momenta of two equal charge pions, $W_2(p,q)$ is the
joint probability of finding one pion with momentum $p$ and one of momentum 
$q$, while $W_1(k)$ is the probability of finding a single pion of
momentum $k$. However, in the OPAL experiment\cite{Ale} as in most other 
experiments the quantity studied is
\begin{equation}
          R_2^{++/+-}(p,q) = W_2(p^+,p^+)/W_1(p^+)W_1(q^-)
\end{equation}
which compares equal charge pion pairs with unequal charge pion pairs.
If all degrees of freedom except the relative momentum\footnote{The momentum 
transfer of two pions with four momenta $p$ and $q$ is 
defined by $Q^2 = M_{pq}^2 - 4 m_{\pi}^2$  where $M_{pq}$ is the 
invariant mass of the pion pair.} $Q$ are summed over one obtains 
\begin{equation}
          R_2^{++/+-} \rightarrow R_2(Q^2)
\end{equation}
which experimentally exhibits a peak at small $Q^2$. This peak is usually 
attributed to the Hanbury-Brown Twiss (H-BT) mechanism\cite{Boa} according to which
pions are emitted with random phases by a source of size $R$. The degree of 
chaoticity of the source is described by the parameter $\lambda$, where 
$0 \le \lambda \le 1$, and $\lambda = 1$ for a fully chaotic source. 

Recently, the enhanced statistics of the OPAL measurements\cite{Ale} at the 
$Z^0$
mass has allowed to study these correlation parameters as a function of 
charged multiplicity $n$, in the range $10 \le n \le 40$. It is found that
the radius $R$ at $n=40$
is about 10 per cent bigger than the radius at $n=10$. At the same time the 
chaoticity parameter decreases by about 15 per cent. These effects, while
small, are statistically significant. We should also mention that detailed 
analysis of H-BT type intensity interference formulae for high multiplicities
\cite{Zai,Cha} shows that the effective radius deduced from experiment tends to 
underestimate the true source radius. For the determination of $R$ and 
$\lambda$, OPAL\cite{Ale} has excluded the contributions from mesons and
resonances decaying into pions, $K^0$, $\omega$, $\eta$, $\eta'$, $\rho^0$,
$f_0(975)$ and 
$f_2(1270)$, at the corresponding high momentum transfers, while no cuts where 
applied below $Q \le 0.3$ $GeV/c$ where the H-BT peak occurs. Due to these 
cuts some
(but not all) dynamical distortions which affect like and unlike charge pion 
pairs differently\cite{Boa,Song} in $R_2$ are excluded and overall the 
chaoticity parameter $\lambda$ becomes smaller than unity. Partly the reduction
in $\lambda$ is also due to experimental acceptance and vertex resolution 
effects. Similar reductions have also been observed at lower 
energies\cite{Ave,Iur}.
In the OPAL analysis\cite{Ale} the increase in source radius for increasing 
pion multiplicity has been linked to a larger fraction of three versus two jet 
events.

In this note we propose that a small increase of the fraction of primary 
mesons different from pions for high pion multiplicities would simultaneously 
explain both the behaviour of $R$ and $\lambda$.  We do not attempt to link our 
mechanism to a more microscopic picture. Our suggestion is presumably not in
conflict with the observation of an increased fraction of three jet events at 
high multiplicities advocated by OPAL\cite{Ale}. Our argument is 
straightforward: Since the observed source radii $R$ in the correlation
function $R_2$ are of order
one fermi any mesons other than pions with long decay paths are introducing 
dynamical elements of coherence by propagating out of the source region. 
The number of primary pions is thereby reduced leading to a reduction of 
$\lambda$, and simultaneously the effective size of the pion source is 
expected to be increasing. Propagation effects have been considered 
previously\cite{Song,Gra,Bow}. We stress that the cuts in the analysis of 
OPAL\cite{Ale} mostly exclude pion correlations from resonance decay
while we now refer to correlations between one primary pion and a pion 
from a decaying meson. In order to produce the subtle effects observed in 
\cite{Ale} it would be necessary that the  f r a c t i o n  of mesons different 
from pions increases for the observed range of multiplicities. We expect 
the strongest effect from fairly long lived mesons having long decay paths, 
like the omega meson for which $\Gamma^{-1}=24$ $fm$. The $\rho$ meson is 
apparently
not in this category, we should, however, remind the reader that no experiments
so far have determined the true pion correlation function $C_2$, but rather the 
quantity $R_2$ has been measured. $R_2$ is a quantity that is much more 
sensitive 
to dynamical correlations, in fact it can be peaked at small relative momenta 
due to different slopes in $Q^2$ for like and unlike charge pion pairs while 
the true correlation function $C_2$ may have no 
peak at all. The fraction of $\rho^0$ 
mesons present in the sample will therefore affect $R_2$ by dynamical depletion
of the unequal charge pion pairs at small $Q^2$, despite of the short $\rho^0$
life time. Bowler\cite{Bow} has shown quantitatively for a chaotic string 
model that propagation effects lead to a larger radius $R$ of the pion 
correlation function if a primary pion is combined with a decay pion from a 
meson resonance. The results depend on the meson resonance in question and on 
the experimental set-up.

Is there experimental evidence for an increase of the fraction of mesons as
a function of multiplicity $n$? A direct experimental determination is not 
available. However, since for a fixed energy the total charged multiplicity is 
sharply peaked and the multiplicity is a well known function of energy,
see Fig.4 in \cite{Dec1} and \cite{Mar,Ade}
it is instructive to inspect the energy dependence of the production of
mesons other than pions. Recent data for electron-positron annihilation
have become very accurate.
In Fig.3 of \cite{Abr} there is clear indication that the
average multiplicity of the vector mesons, $\rho^0$ and $K^{\star}$, per 
hadronic event, increases with energy. Denoting the average $\rho$ meson 
multiplicity by $n(\rho)$ and the average 
charged multiplicity by $n$, as before, we read from the quoted
figures $n(\rho)=0.8$, $n=13$ at $30$ $GeV$, and $n(\rho)=1.4$, $n=20$ 
at $90$ $GeV$. Similar
results hold for the fairly narrow $K^{\star}$. It is plausible that other 
mesons ($\omega$, $K$, $\eta$, $\eta$', etc.) show similar trends. Such 
behaviour could therefore partially explain the observed multiplicity 
dependence in \cite{Ale}. Indeed if one
calculates straightforwardly from the numbers given above the fractional 
population of $\rho$'s at $30$ and $90$ $GeV$ one can also estimate the 
derivative of
this fraction with respect to multiplicity $n$ to be roughly 1 per cent and 
POSITIVE. This value is slightly large when compared with the slope for the 
radius parameter $R$ from OPAL\cite{Ale} which is 
$1/\langle R\rangle \star dR/dn = (3.6 \pm 0.6) \times 10^{-3}$
for the Goldhaber parametrization. However, we must keep in mind that most of 
the propagation effects happen at very small momentum transfers, cp.
Bowler\cite{Bow}. In the OPAL experiment\cite{Ale} the relative momenta are 
restricted to 
$Q^2_{exp} > (0.05 GeV)^2$, and the effect is therefore expected to be diluted 
by the cut in $Q^2$ and by losses due to vertex resolution\cite{Act3}. 
In other reactions,
like multiple pion production in heavy ion collisions\cite{Bol,Alb}, similar 
results are to be expected. Finally we would like to stress that the 
mechanisms suggested here for explaining the multiplicity dependence of the 
$R$ and $\lambda$ parameters do not 
necessarily depend on the stochastic picture for the pion emission region. 
All that is needed here is a nontrivial correlation function for the primary 
pion emission, be it $C_2$ or $R_2$. Nontrivial correlation functions can 
also be obtained by averaging procedures different from the stochastic H-BT 
model by various summations over unobserved variables which involve an 
entirely coherent basic pion-emission process\cite{Boa,Song,NAN,Ama1,Ama2} 
from more than one point source.
From a theoretical point of view resonance effects can be treated in the 
Skyrmion picture\cite{Yan} combined with the coherent state formulation 
of pion emission\cite{Ama1,Ama2}, there may also be an 
interesting connection to squeezed coherent states\cite{Wei,And,Dre}.

  We would like to thank Valeri Markushin for discussions.

\end{document}